\documentclass[intlimits,twoside,a4paper]{article}

\usepackage{graphicx}        %%%% already in cmpj3.sty

\usepackage[T2A,T1]{fontenc} %%%% already in cmpj3.sty
\usepackage[cp1251]{inputenc}

\usepackage[eqsecnum]{cmpj3}
\issue{2021}{24}{2}{23704}
\doinumber{10.5488/CMP.24.23704}

\title[One-particle spectral densities]%
{One-particle spectral densities and phase diagrams of
one-dimensional proton conductors}

\author[R. Ya. Stetsiv]
{R. Ya. Stetsiv
\thanks{\email{stetsiv@icmp.lviv.ua}}}

\address{Institute for Condensed Matter Physics of the National
Academy of Sciences of Ukraine,\\ 1 Svientsitskii St., 79011 Lviv,
Ukraine
}

\date{Received November  19, 2020, in final form April 2, 2021}
\begin{document}

\maketitle

\begin{abstract}
The equilibrium states of one-dimensional proton conductors in the
systems with hydrogen bonds are investigated. Our extended hard-core
boson lattice model includes short-range interactions between
hydrogen ions, their transfer along the hydrogen bonds with
two-minima local anharmonic potential, as well as their inter-bond
hopping, and the modulating field is taken into account.  The exact
diagonalization method for finite one-dimensional system with
periodic boundary conditions is used. The  existence of various
phases of the system at $T = 0$, depending on the values of
short-range interactions between particles and the modulating field
strength, is established by analyzing the character of the obtained
frequency dependence of one-particle spectral density; the phase
diagrams are built.
%
%\keywords proton conductor, hard-core boson model, spectral density,
%phase diagrams
%
%\pacs 75.10.Pq, 03.75.Lm, 66.30.Dn
\printkeywords
\end{abstract}

\section{Introduction}

Ionic conductors are known since the times of Faraday
\cite{Faraday}, but the biggest urge for the development of this
direction in solid state physics was the discovery in 1960th of
the  structures with charge carriers that are ions of silver ($\alpha$ ---
AgI, Ag$_{2}$S). These crystals are characterised by the  presence of
high temperature phase with high ionic conductivity. This phase
afterwards was referred to as superionic (the same as crystals). The
further research resulted in the opening of a new class of the
systems where charge carriers are hydrogen ions. They are
ferroelectric or ferroelastic crystals at low temperatures, but at
higher temperatures they undergo transition to superprotonic phase,
while the conductivity is increased by several orders of magnitude
(among others there are compounds of the general form MeHXO$_{4}$,
where Me = Cs, Rb, NH$_{4}$;  X = S, Se). Numerous structural studies
have shown that in low-temperature phase the protons are clearly in
the fixed positions, while in high-temperature phase they are
distributed with equal probability between multiple positions in the
unit cell. Much effort is presently put for the synthesis of new
structures with high ionic conductivity. This is predetermined by
their use in the areas of new technique such as hydrogen energetics,
chemical sources of current, electronics, control and measurable
devices for special purposes.  As an example, we can cite a series
of lithium conductive materials synthesized from perovskite
structures La$_{23-x}$Li$_{3x}$TiO$_{3}$-type
\cite{Belous1,Belous2,Belous3}. The superionic state is educed, for
example, in Ag$_{2}$S,  AgI, AgBr, CuBr, Cu$_{2}$S, CuCl, RbAg$_{4}$I$_{5}$
compounds where the metallic cation migrates. Structures with the
large concentration of admixture ions: oxidizing solid solutions
MO$_{2}$-M$'_{2}$O$_{3}$ and MO$_{2}-$M$''$O type, where M --- Zr, Ge; M$'$ ---
Ca, Sr, Ba; M$''$ --- S, Y; structures of Na$_{2}$O$\,\cdot11$Al$_{2}$O$_{3}$-type
(Na migrates in planes between Al$_{2}$O$_{3}$ blocks). Without
regard to great successes, only a few substances with high ionic
conductivity and stable against chemical and mechanical action and
possessing other specific properties are presently known. The
conductivity of ionic conductors is particularly high when the
number of ions is much less than the number of positions in a
lattice, i.e., when there are vacancies. Therefore, a lot of free
positions facilitate the probability of ion hopping from one
position to another. In particular, in the superionic phase of AgI
($T>147^{\circ}$C), $2$ conductivity ions of Ag$^{+}$ are statistically
distributed between 42 settled positions of different type in the
unit cell.

Models of a lattice gas type with Pauli statistics of particles are often used for the description of ionic
conductors \cite{Mahan,Stasyuk5,Stasyuk6,Stetsiv,Micnas1,Micnas2,Rigol,Rousseau,Rigol1,Rigol2,Hen2,Batrouni1,Batrouni2,Bernardet1,Bernardet2,Schmid}.
Here, particles are of Bose nature but they also obey the Fermi rule. The lattice model of Pauli particles is
similar to the Bose-Hubbard model in the hard-core approximation (provided that the occupation numbers are
restricted, $n_{i}$ = $0.1$). Such a lattice model can describe the emergence of a superfluid (SF)-type state
even in the absence of a direct interaction between particles \cite{Stasyuk5,Stasyuk6,Stetsiv}.

One-dimensional ion conductors were investigated in our previous
work \cite{Stetsiv}. The existence of various phases of the system,
depending on the values of interactions between particles $V$ and
modulating field strength $A$, was established. In this work we
examine finite one-dimensional ionic conductors  with the two-minima
local potential for ions. Usually, they are proton conductors in the
systems with hydrogen bonds. $\emph{Ab initio}$ studies of many
researchers show that in the systems with hydrogen bonds, there
remains $0.75 - 0.8$ of electron ($0.75e - 0.8e$) at the hydrogen ion.
Despite this, it is generally accepted  to call conductors with
hydrogen bonds proton conductors. We have got not a proton but an
ion of hydrogen having a charge $+0.2\div+0.25$. Thus, we cannot
state that these are Fermi-particles or Bose-particles, so we use
hard-core boson lattice model. Our extended hard-core boson lattice
model includes short-range interactions between ions, their transfer
along hydrogen bonds with the two-minima local anharmonic potential
as well as their inter-bond hopping, and modulating field is taken
into account. Here, unlike the previous case, we have two different
interactions between ions ($V$ on the hydrogen bond, and \textit{w} between
bonds) and consequently two different transfers. An exact
diagonalization method for a finite one-dimensional system with
periodic boundary conditions is used. Energy spectrum and
one-particle spectral densities are calculated; the phase diagrams
are built.

\section{The model}

We use the lattice gas quantum model for the description of ionic
(proton) conductors. This model corresponds to the hard-core boson
limit, if the particles obey the Pauli statistics. If the  positions
of particles are characterized by symmetric local  potential
possessing two minima on the bond, the two stage nature of particle
transport is taken into account via two constants of particle
transfer ($\Omega_0$ for intra-bond hopping between two positions
$a$ and $b$ on each bond as well as $\Omega_R$ for inter-bond
transfer between hydrogen bonds that arises due to orientational
motion of ionic groups). That is why the model is referred to as
orientational-tunneling model \cite{Stas97}. It also includes the
correlation between nearest ions caused by the short-range repulsion
(the corresponding energies parameters are: $V$ on the bond and \textit{w}
for the neighbor bonds); the modulating field is also included (the
parameter $A$). The field $A$ causes a spatial modulation of the ion
distribution in the so-called ordered phase (the existence of such a
phase at low temperatures is a characteristic feature of superionic
conductors).

The Hamiltonian of this model for the chain-like structure  can be
presented as follows:
\begin{eqnarray}
    H
    &=&
    -\mu \sum\limits_{i} (n_{ia}+ n_{ib} ) + V \sum\limits_{i} n_{ia} n_{ib}
    + \textit{w} \sum\limits_{i} n_{ib} n_{i+1,a} + \Omega_{0} \sum\limits_{i} (c_{ia}^{+}c_{ib} + c_{ib}^{+} c_{ia}) 
    \nonumber
    \\
    &+&\Omega_{R} \sum\limits_{i} (c_{ib}^{+} c_{i+1,a} + c_{i+1,a}^{+} c_{ib}) + A \sum\limits_{i} (n_{ib} - n_{ia})\,.
\label{ham}
\end{eqnarray}
Here, operators $c_{i,\alpha}$ ($c^+_{i,\alpha}$)  are the hard-core
boson (HCB) operators. They describe the process of annihilation
(creation) of particle on bond $i$ in position $\alpha$
($\alpha=a,b$); therefore, $n_{i,\alpha}=c^+_{i,\alpha}
c^{\phantom{+}}_{i,\alpha}$ is the occupation number of particles in this position
(here, the eigenvalues of $n_{i,a}$ and $n_{i,b}$ are equal to $0$
or $1$). At different sites, HCB creation and annihilation operators
commute as usual of bosons:
\begin{eqnarray}
[c_{k},c^+_{m}] = [c_{k},c_{m}] = [c^+_{k},c^+_{m}] = 0,\quad k\neq m\,.
\end{eqnarray}
At the same site these operators satisfy anticommutation relations
typical of fermions:
\begin{eqnarray}
\{c_{m},c^+_{m}\} = 1, \quad c^{+2}_{m} = c^2_{m} = 0.
\end{eqnarray}
 The energy spectrum of finite one-dimensional
system with periodic boundary conditions is  calculated using the
exact diagonalization method. The Hamiltonian matrix as well as
$c_{i,a}$ and $c^+_{i,a}$ matrices are constructed on the basis of
many-particle states $| n_{1,a} n_{1,b} \ldots n_{N,a} n_{N,b}
\rangle$. The Hamiltonian matrix on the basis of these states is
diagonalized numerically. Such an operation corresponds to the
transformation
\begin{eqnarray}
U^{-1} H U = \widetilde{H} = \sum \limits_p \lambda_p
\widetilde{X}^{pp},
\end{eqnarray}
where $\lambda_p$ are eigenvalues of the Hamiltonian,
$\widetilde{X}^{pp}$ are Hubbard operators (in general,
$\widetilde{X}^{pq} = \mid p \rangle\langle q\mid$), (see
\cite{Hubbard65}, also \cite{Stas13a}). The creation and
annihilation operators are presented in the form
\begin{equation}
U^{-1} c_{i,\alpha} U = \sum \limits_{pq} A^{i,\alpha}_{pq}
\widetilde{X}^{pq}\,, \quad \quad U^{-1} c^+_{i,\alpha} U = \sum
\limits_{rs} A^{i,\alpha*}_{rs} \widetilde{X}^{sr},
\end{equation}
where coefficients $A^{i,\alpha}_{pq}$ are the matrix elements of
operator $c_{i,\alpha}$ on the new bases.

Like in our previous work \cite{Stetsiv}, we construct two-time
temperature Green's functions $G_{i,\alpha;i,\alpha} =
\langle\langle c_{i,\alpha} | c^+_{i,\alpha} \rangle\rangle$
containing the information about the one-particle spectrum of the
system. We introduce Green's functions of two types, namely, the
commutator Green's function
\begin{eqnarray}
\langle\langle c_{i,\alpha} (t) | c^+_{i,\alpha} (t') \rangle\rangle
^{(c)} = - \ri \Theta (t-t') \langle [c_{i,\alpha} (t), c^+_{i,\alpha}
(t')] \rangle \label{grcom}
\end{eqnarray}
and the anticommutator Green's function
\begin{eqnarray}
\langle\langle c_{i,\alpha} (t) | c^+_{i,\alpha} (t') \rangle\rangle
^{(a)} = - \ri \Theta (t-t') \langle \{ c_{i,\alpha} (t),
c^+_{i,\alpha} (t') \} \rangle. \label{grant}
\end{eqnarray}
One-particle spectral densities are determined by the imaginary
parts of those  Green's functions
\begin{eqnarray}
 \rho(\omega) &=& - \frac{1}{\piup N} \sum \limits_{j=1}^N \sum
\limits_{\alpha} \textrm{Im} \langle\langle c_{j,\alpha} |
c^+_{j,\alpha} \rangle\rangle_{\omega  + \ri\varepsilon}
\label{greenf}
\\ &=& - \frac{1}{\piup N} \sum \limits_{j=1}^N \sum
\limits_{\alpha} \textrm{Im} \left[\frac{1}{Z} \sum \limits_{pq}
A^{j,\alpha}_{pq} A^{j,\alpha*}_{pq} \frac{\re^{-\beta \lambda_p}
\!-\! \eta \re^{-\beta \lambda_q}}{\omega \!-\!
\frac{1}{\hbar} (\lambda_{q}\!-\!\lambda_{p}) \!+\! \ri\varepsilon} \right]. \nonumber
\end{eqnarray}
Here, $Z = \sum \limits_p \re^{-\beta \lambda_p}$ .

We obtain spectral densities of commutator Green's function
(\ref{grcom}), when $\eta = 1$, and anticommutator Green's function
(\ref{grant}), when $\eta = -1$. Spectral densities have a discrete
structure that includes a number of $\delta$-peaks owing to the
finite chain size. If the chain size (i.e., the number of sites $N$)
increases, the $\delta$-peaks are located more densely and, at $N
\rightarrow \infty$, they form a band structure. We confined
ourselves to the case $N = 12$. We introduce the small parameter
$\Delta$ to broaden the $\delta$-peaks in accordance with Lorentz
distribution $\delta (\hbar\omega) \rightarrow \frac{1}{\piup}
\frac{\Delta}{(\hbar\omega)^2+\Delta^2}$.

\section{One-particle spectral densities and phase diagrams}

In this work all calculations are performed for the temperature
equal to zero ($T=0$). Numerical values of all energy parameters
(including $\hbar \omega$) are presented in units of parameter
$\Omega_{0}$, and it is dimensionless.  Experimental data,
quantum-chemical calculations, semiempiric theoretical estimations
offer a wide region of values of interaction between ions, $V =
3\cdot10^{3}...10^{4}$ cm$^{-1}$, $\textit{w} = 10^{3}...10^{4}$ cm$^{-1}$,
depending on the objects that are examined
\cite{Munch95,Hassan92,Eckert}. The ion transfer parameter
$\Omega_{0}, \Omega_{R}$ can vary within wide limits,
$40\ldots2500$ cm$^{-1}$. For example, it is obtained $V =
5\cdot10^{3}\ldots10^{4}$ cm$^{-1}$ from the experimental data for $T_c$
in the case of $H$-bonded ferroelectrics. In our calculations we
chose $\Omega_{R}/\Omega_{0} = 0.5$. We changed the parameters of
short-range interactions in wide limits: $V/\Omega_{0} =
0,1,\ldots,10$, $\textit{w}/\Omega_{0} = 0,1,\ldots,10$. It is necessary to note
that the problem is invariant in relation to simultaneous
replacement of numerical values  $\textit{w} \leftrightarrow V,\Omega_{R}
\leftrightarrow \Omega_{0}$. For convenience, we use the notation
$\mu' = \mu - (V+\textit{w})/2$.

One-particle spectral densities are calculated according to formula
(\ref{greenf}). The  existence of various phases of the system at $T
= 0$, depending on the values of interactions between particles  and
the modulating field strength, is established by analyzing the
character of frequency dependence of one-particle spectral density.
According to works \cite{Menotti,Stasyuk13b}, a characteristic
feature of the commutator spectral density in the superfluid (SF)
phase is  the continuous continuation  at $\omega = 0$ of a negative
branch (which exists at $\omega < 0$) to a positive branch (which
exists at $\omega > 0$). The chemical potential of particles is
located at the point $\omega = 0$. In the charge ordering (CDW)
phase, these branches are separated by the gap. The chemical
potential of particles is located in the energy gap. We obtain a
split of the spectrum into two subbands and the emergence of a
modulated state. Charge-density-wave (CDW) state is the
characteristic of the case of a half-filling of ionic sites
($\langle n \rangle = 1/2$) and one may observe the situation when
all protons occupy ``$a$'' positions (or all protons occupy ``$b$''
positions) along the chain. This case corresponds to ferroelectric
type ordering, though it has a more general meaning. The protons
occupy only some of the positions available (while other positions
remain unoccupied) which is a general feature of the ordered phases
that exist in superionic crystals. We call this state CDW though the
doubling of the lattice period is not observed. For the case of
ionic conductor (with one minimum local potential for ions), the
splitting of spectra occurred due to the charge ordering with the
doubling of lattice period (see \cite{Stetsiv}). If one attempts to
include the long-range interaction
 to our model, he or she will get the doubling of the lattice period also for proton conductor as well.
 At moving away from the half-filling, we
get into the superfluid (SF) state. In this phase, the
conductivity of the system grows by a few orders.  Such a state
experimentally looked like a superionic phase. As we go far away
from half filling (for example, a decrease of chemical potential
$\mu$) we get into a Mott insulator (MI) state. In this state, the
commutator spectral density has no negative branch. The chemical
potential is below the band and ions need some activation energy to
induce their transport. MI state can also be observed when the
chemical potential is located above the upper subband, and the
commutator spectral density has only a negative branch. The average
occupation number of the state at a given $\mu$ is calculated
according to the spectral theorem, $\langle n \rangle =
\int_{-\infty}^{\infty} \frac{\rho_a (\omega) \mathrm{d}
\omega}{\mathrm{e}^{\beta\omega} + 1}$, where $\rho_a$ is the
anticommutator spectral density (the density of states).

\begin{figure}[!b]
\centering\includegraphics[width=0.60\columnwidth]{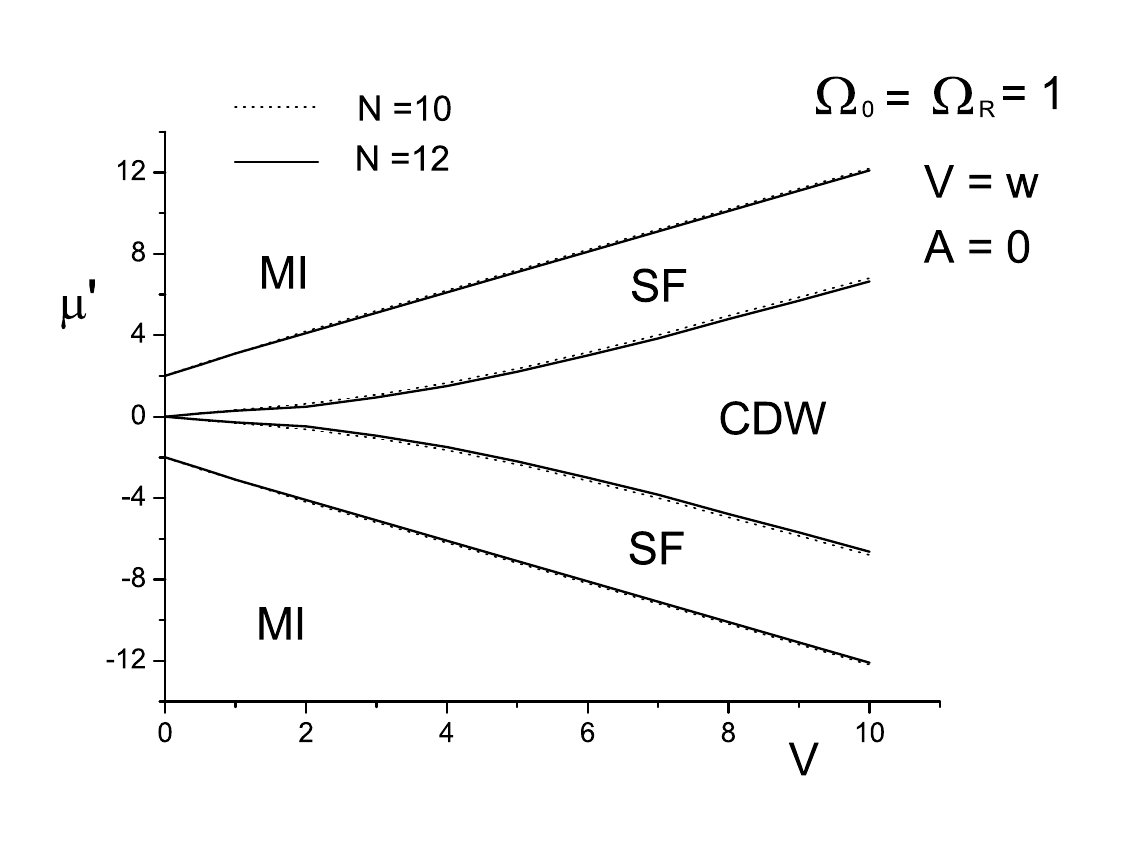}
%\vskip-5mm 
\caption{Phase diagram for a one-dimensional ionic
conductor in the $(\mu', V)$ coordinates, $(T = 0)$.} 
\label{fig1}
\end{figure}

\begin{figure}[!t]
\centering\includegraphics[width=0.60\columnwidth]{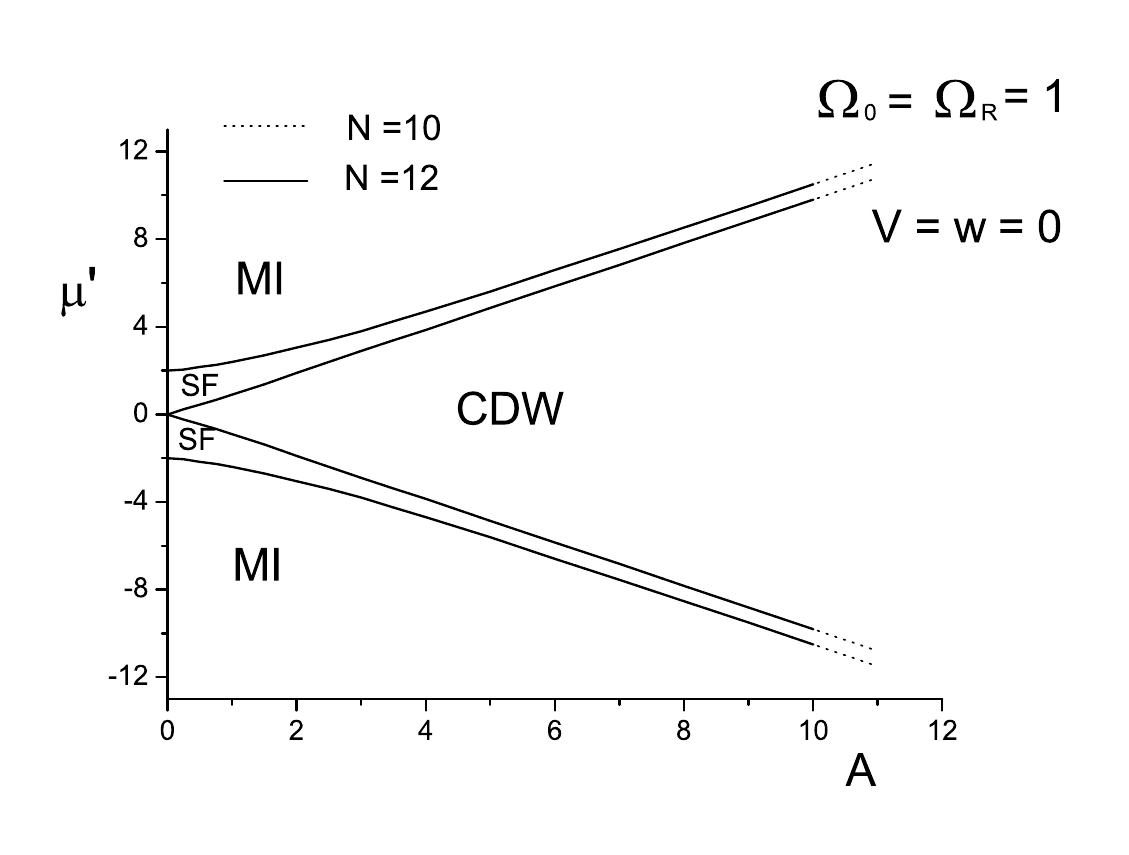}
%\vskip-5mm 
\caption{Phase diagram for a one-dimensional ionic
conductor in the $(\mu', A)$ coordinates, $(T = 0)$.} \label{fig2}
\end{figure}

In previous work \cite{Stetsiv}, there  was examined a chain with
ten positions for  ions ($N = 10$). For comparison, we repeated some
previous calculations for the one dimensional systems with the
one-minima local potential for the conductivity particles, though at
$N = 12$ , in order to see the dependence of the results on~$N$. The
Hamiltonian of this problem is received from the Hamiltonian for the
systems with two-well local anharmonic potential for ions
(\ref{ham}), provided $V =\textit{w}$  and $\Omega_{R} = \Omega_{0}\equiv t$.
In figure \ref{fig1} and figure \ref{fig2} some diagrams of state at $N
= 12$ are compared with the ones obtained earlier in \cite{Stetsiv}
at $N = 10$.
For the case, with the  dependence of the phase  borders on the
value of interaction between particles $V$ shown in figure \ref{fig1},
which is a phase diagram in the ($\mu',V$) coordinates, the
numerical difference in the position of the phase-to-phase border
(in the $\mu'$ coordinate) between the cases of $N = 10$ and $N =
12$ is 2--3 percent. For the case of phase diagram in the
($\mu',A$) coordinates, (figure \ref{fig2}), this difference is $0.5$
percent. The last one is built for the case $V = 0$ and coincides
with an exact diagram obtained analytically in the works
(\cite{Rousseau,Rigol1,Rigol2,Rigol}, see also \cite{Hen2}). An
exact analytical solution can be obtained by applying the
Jordan-Wigner transformation, which makes it possible to pass from
the Hamiltonian of hard-core bosons to the Hamiltonian of
noninteracting spinless fermions (only in one-dimensional case).
 In this diagram, the lines separating the CDW and SF phases are linear on the field $A$ and look like $\mu' = +A$ and $\mu' = -A$ (see figure \ref{fig2}).

In this work we examine a finite one-dimensional ionic conductors
with the two-minima local anharmonic potential for ions. Usually
these are proton conductors in the systems with hydrogen bonds. All
calculations are executed for $N$ = 12.

We obtained the phase diagrams of equilibrium states of the system
depending on the interactions between ions and modulating field.

\begin{figure}[!t]
\centering\includegraphics[width=0.60\columnwidth]{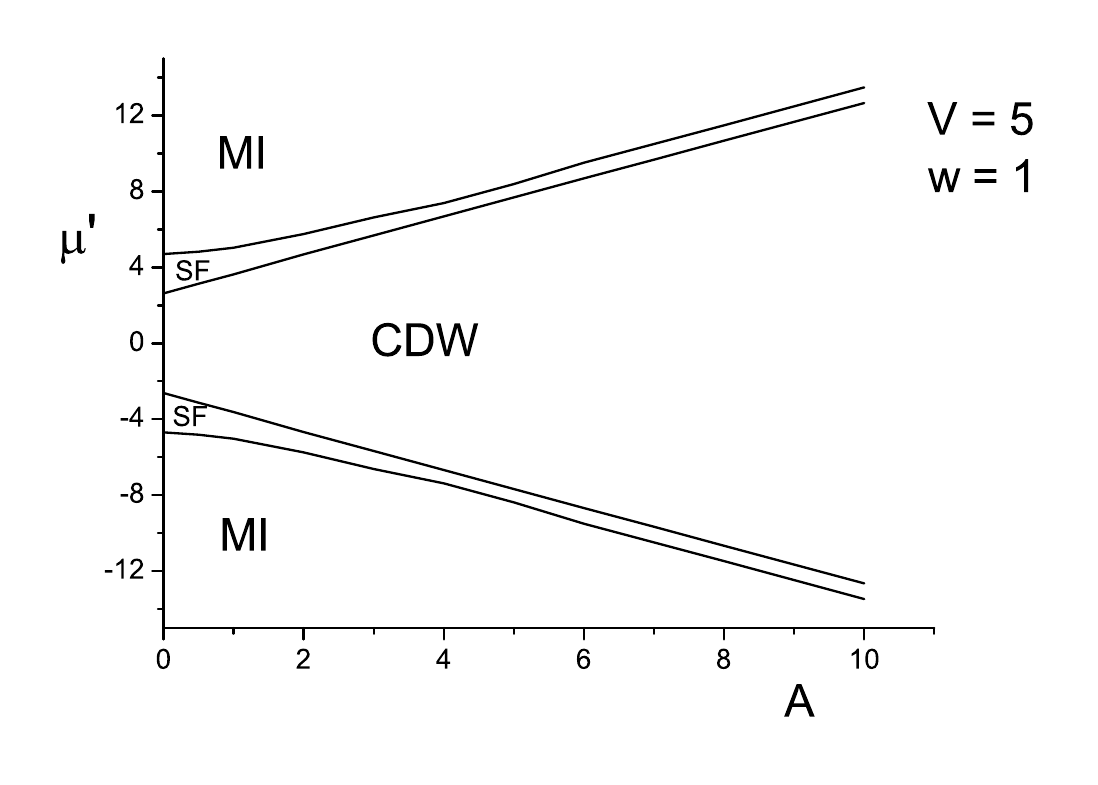}
\caption{Phase diagram for a one-dimensional proton
conductor in the $(\mu', A)$ coordinates, $(T = 0)$.} \label{fig3}
\end{figure}

For example, the diagram of state showing the dependence on the
modulating field $A$ is presented in figure \ref{fig3} (here $V = 5$, $\textit{w} = 1$). It is a diagram in the ($\mu',A$) coordinates.
The line separating CDW and SF phases in coordinates ($\mu',A$)
is a straight line depending on $A$ (see figure \ref{fig3}). The width
of CDW phase (in $\mu'$ coordinates) grows with an increase of the
value of the modulating field.

\begin{figure}[!t]
\begin{minipage}[h]{0.47\linewidth}
\center{\includegraphics[width=1\linewidth]{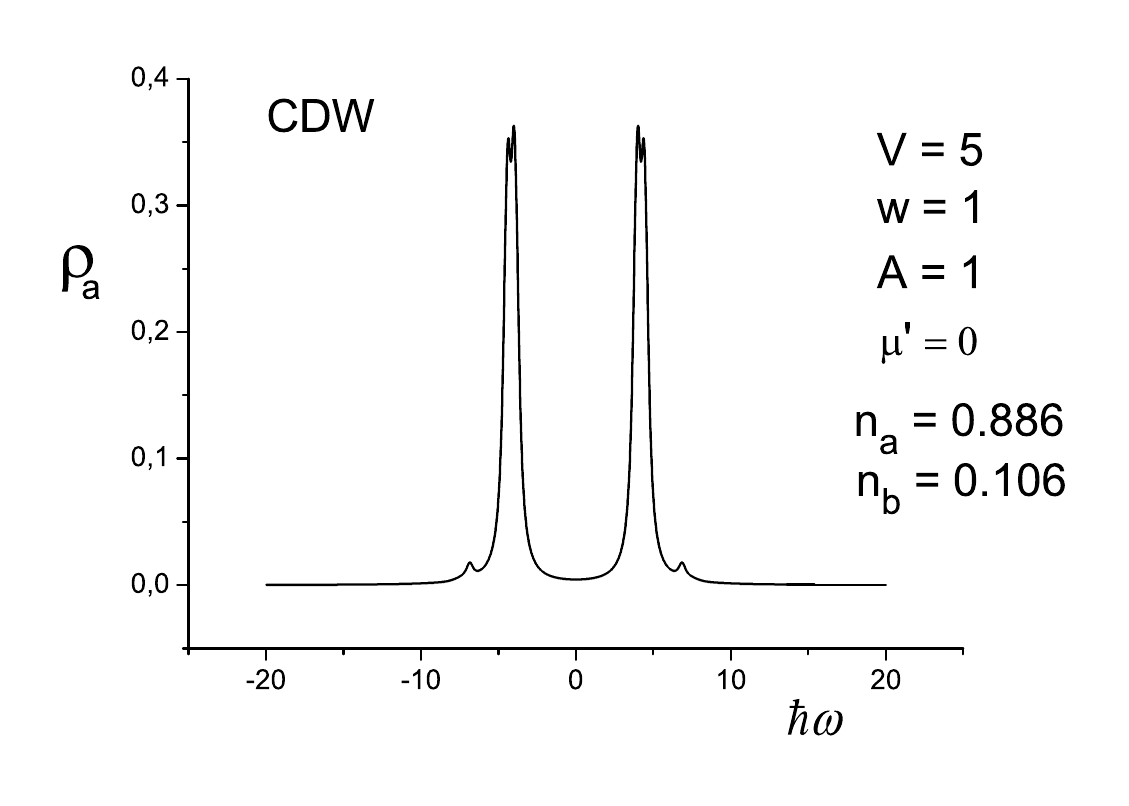}} (a) \\
\end{minipage}
\hfill
\begin{minipage}[h]{0.47\linewidth}
\center{\includegraphics[width=1\linewidth]{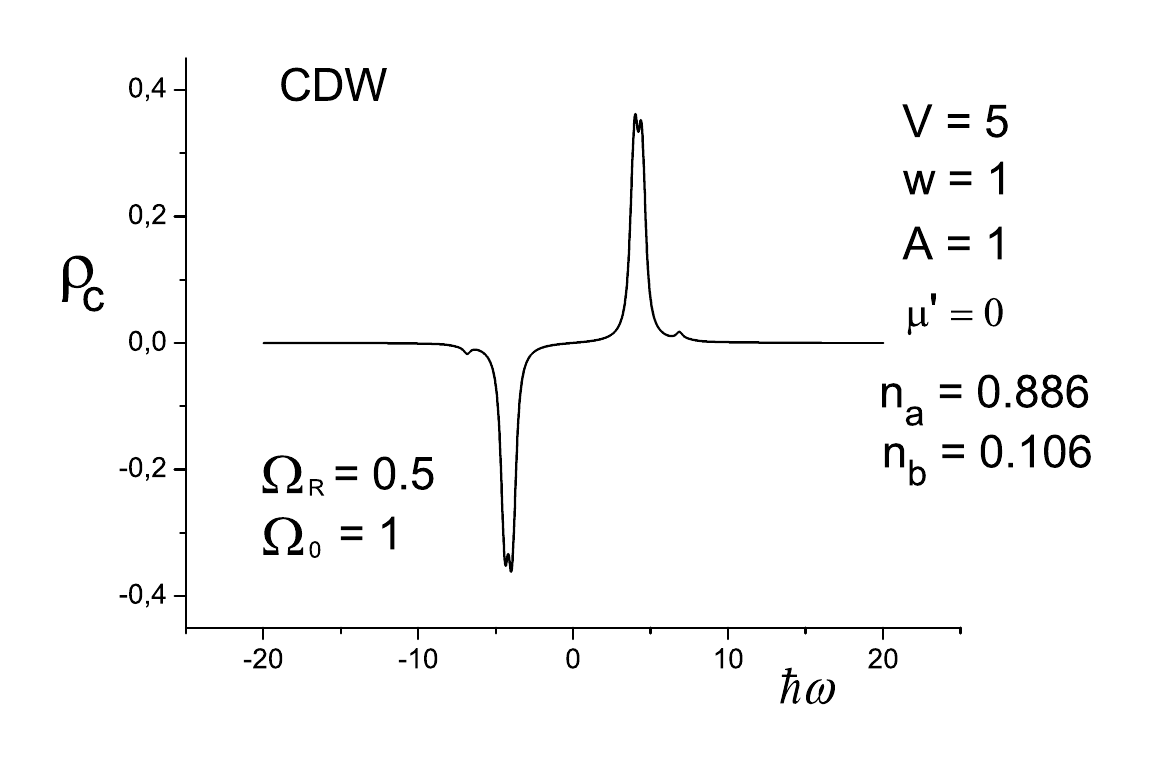}} \\(b)
\end{minipage}
\vfill
\begin{minipage}[h]{0.47\linewidth}
\center{\includegraphics[width=1\linewidth]{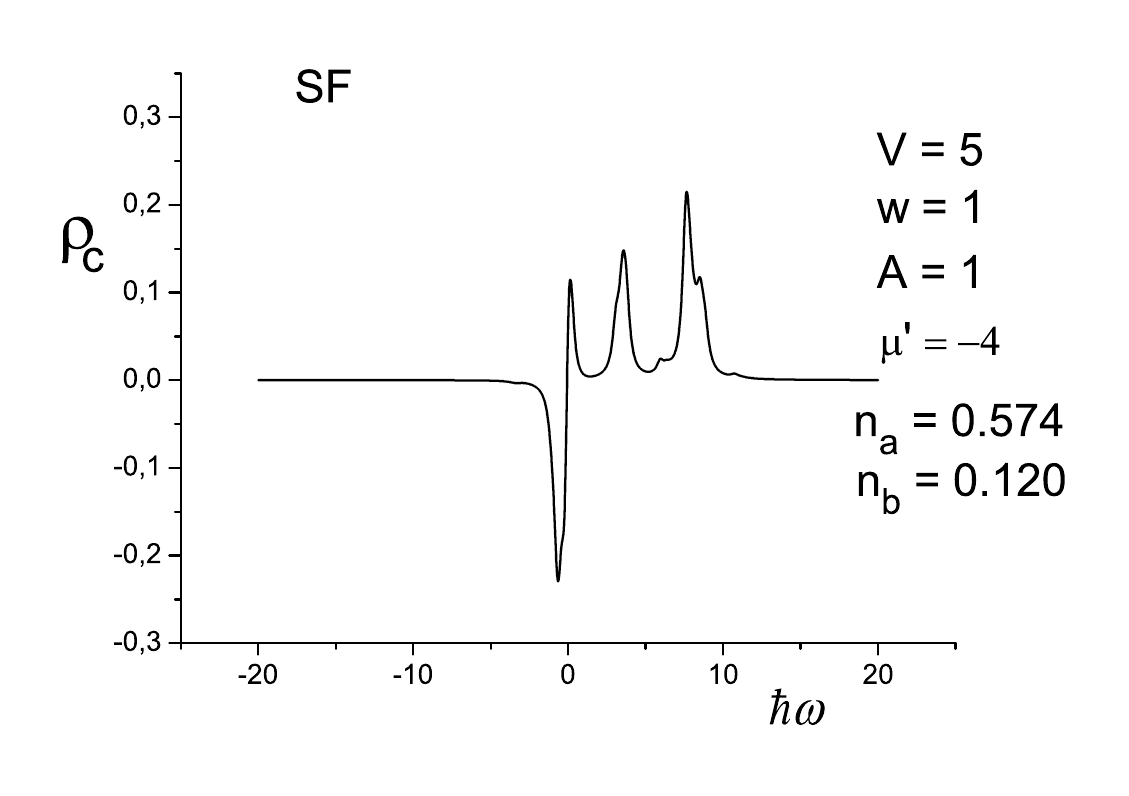}} (c) \\
\end{minipage}
\hfill
\begin{minipage}[h]{0.47\linewidth}
\center{\includegraphics[width=1\linewidth]{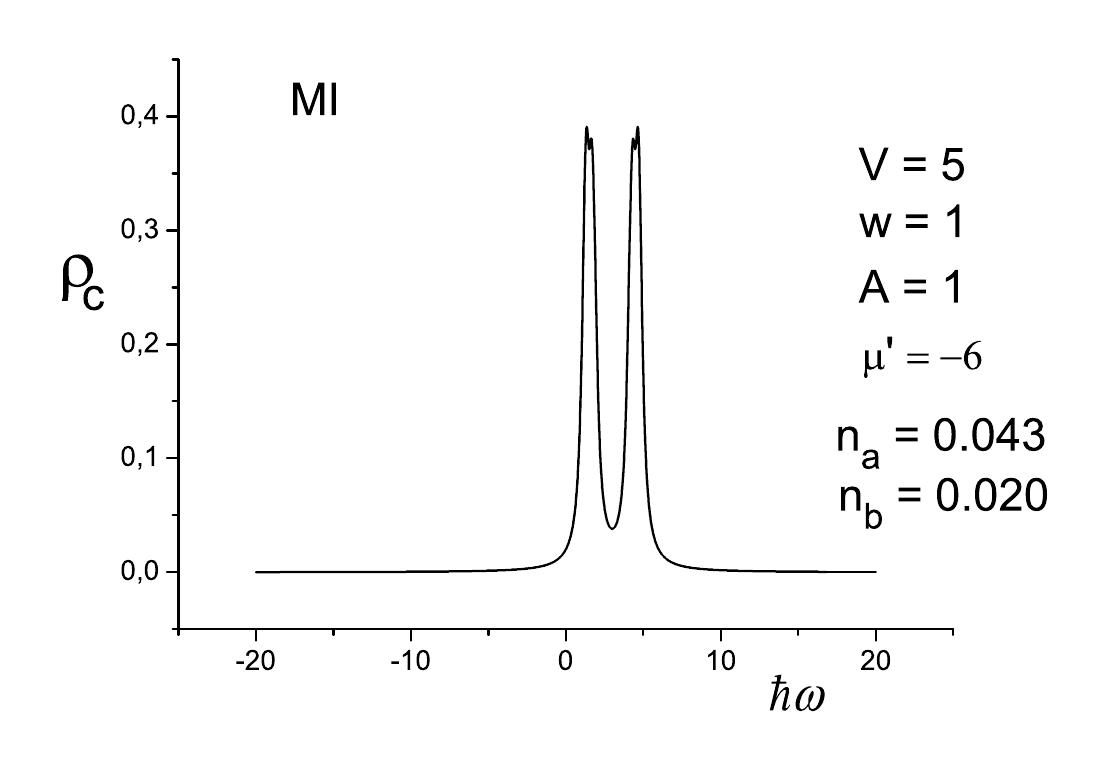}} (d) \\
\end{minipage}
\caption{Anticommutator (a) and  commutator (b-d) one-particle
spectral density for various states of a one-dimensional proton
conductor; $V = 5$, $\textit{w} = 1$, $A = 1$, $T = 0$, $\Delta = 0.25$. The chemical
potential level is located at $\omega = 0$.}
\label{fig4}
\end{figure}

Anticommutator and commutator one-particle spectral densities, that
correspond to this diagram at $A = 1$, are shown in figure \ref{fig4}.
We move from one phase to another by changing the chemical
potential. The average occupation number of positions ``$a$'' and
``$b$'' in two-minima potential for each phase
 is presented in figure \ref{fig4}. For example, in CDW phase (at $A = 1$), we obtained $n_{a} = 0.886$, $n_{b} = 0.106$,
protons occupy mainly ``$a$'' positions in two-minima local potential on the bond.
 The width of CDW phase is determined by the width of a gap in the energy spectrum.
%
% % % % % % % % % % % %
% % % % % % % % % % % % %

Chemical potential level is placed at $\omega = 0$. In CDW phase,
chemical potential level is in the gap, while in SF phase we
observe a continuous transformation of the negative branch of the
commutator spectral density into the positive one at $\omega = 0$.
\begin{figure}[!t]
\centering\includegraphics[width=0.55\columnwidth]{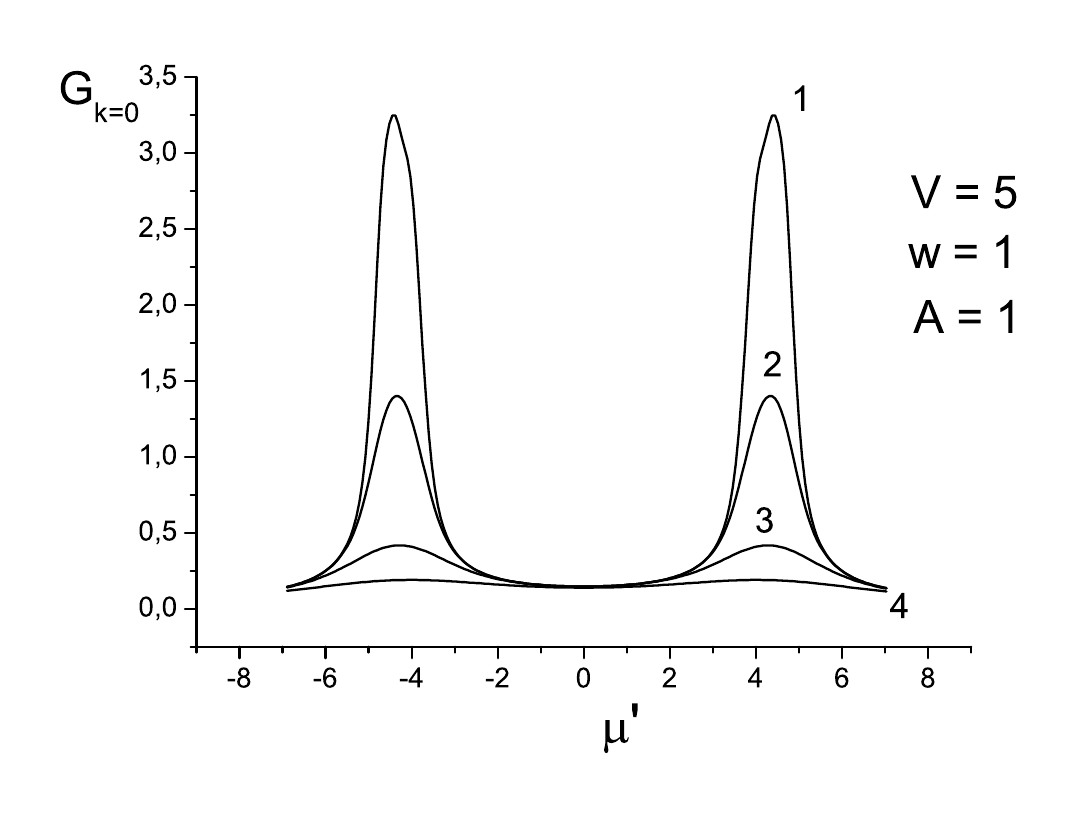}
%\vskip-5mm
\caption{Fourier transform of the real part of commutator Green's
function at zero frequency ($\omega = 0$) and zero wave vector $(k
= 0)$, $\Re G_{k=0}(\omega = 0)$: $1 - T = 0.1$; $2 - T =  0.2$; $3 - T = 0.5$;
$4 - T = 1$; $\Delta = 1\times10^{-6}$.} 
\label{fig5}
\end{figure}

In work \cite{Stasyuk9} it is shown that an important characteristic of the SF phase is a divergence of the
Fourier transform of the real part of commutator Green's function at zero frequency ($\omega = 0$) and zero wave vector
$(k = 0)$, $\Re G_{k=0}(\omega = 0) \rightarrow \infty$. In the case of the considered finite-chain model
\begin{eqnarray}
&& G_{k=0}(\omega = 0) = \frac{1}{N} \sum \limits_{i=1}^N \sum
\limits_{j=1}^N \left[\frac{1}{Z} \sum \limits_{pq} A^{i}_{pq}
A^{j*}_{pq} \frac{\re^{-\beta \lambda_p} - \re^{-\beta
\lambda_q}}{\lambda_{p} - \lambda_{q} + \ri\varepsilon} \right].
\label{gf}
\end{eqnarray}
The static susceptibility $\Re G_{k=0}(\omega = 0)$ calculated by us
arrives at maximal values in SF phase,
 though even at $T=0$ those values remain finite, which is a result of the finite size of the chain
and probably with unidimensionality of the object.
 $\Re G_{k=0}(\omega = 0)$ at $V = 5$, $\textit{w}= 1$, $A = 1$ is shown in figure~\ref{fig5} ($T \neq 0$).
At $T=0$ (at the same parameters) we get a sharp peak in the region
of SF phase, and susceptibility reaches the value of $1778.6$,
which is by three orders larger than at $T=0.1$ (see
figure~\ref{fig5}). At the further increase of temperature, a maximum
$\Re G_{k=0}(\omega = 0)$ becomes smeared, which proves that in one-dimensional system a SF phase exists only at $T = 0$. The regions
of different phases at the above-mentioned parameters can be seen on
the state diagram in figure \ref{fig3}
% % % % % % % % % % % %
\begin{figure}[!b]
	\centering
	\includegraphics[width=0.45\linewidth]{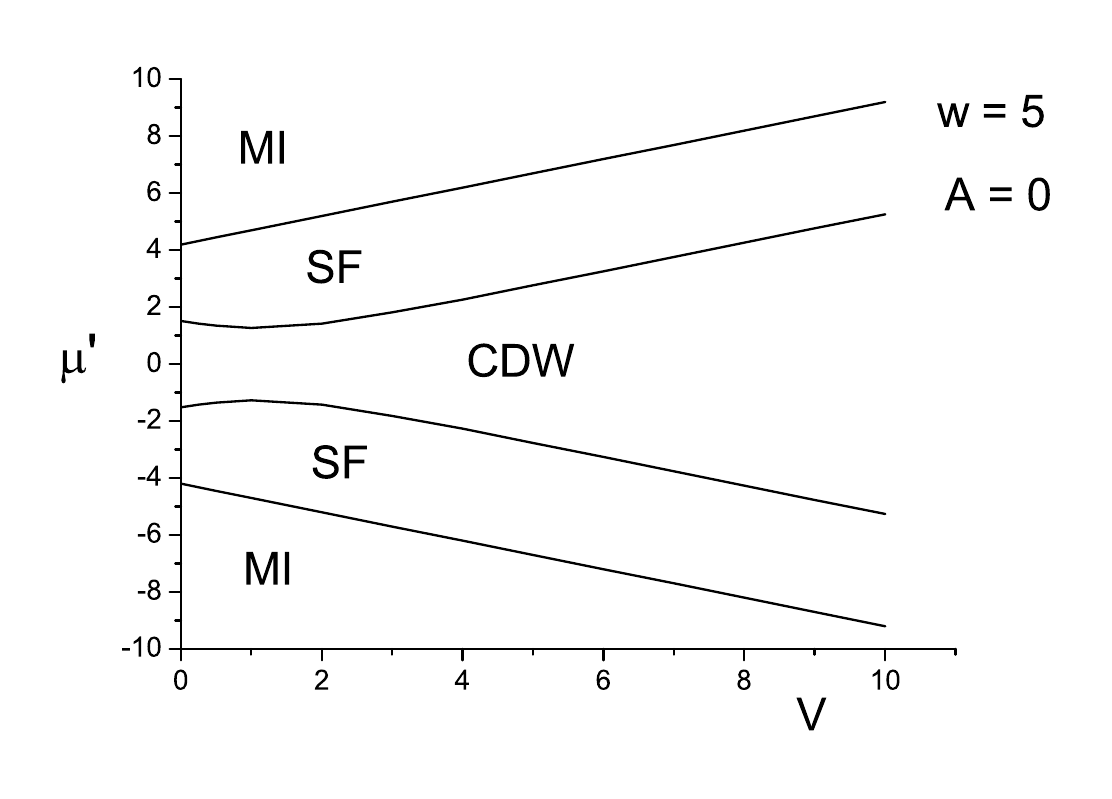} \qquad
	\includegraphics[width=0.45\linewidth]{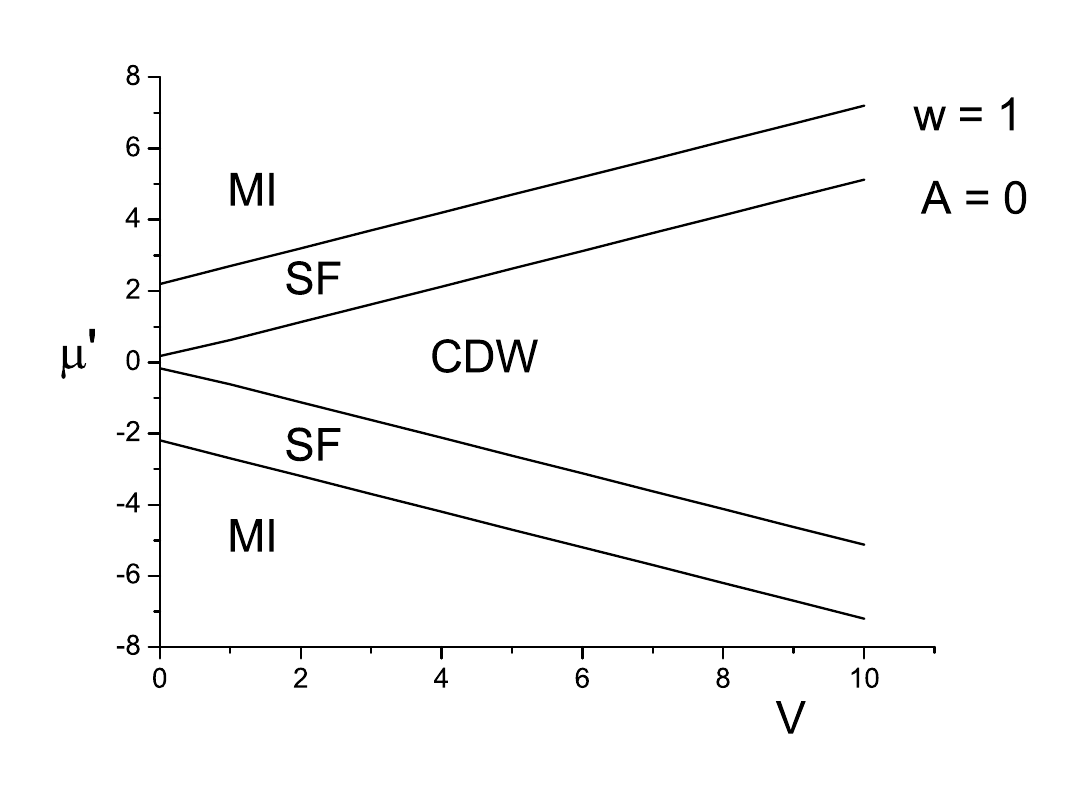} \\%\quad
	\includegraphics[width=0.45\linewidth]{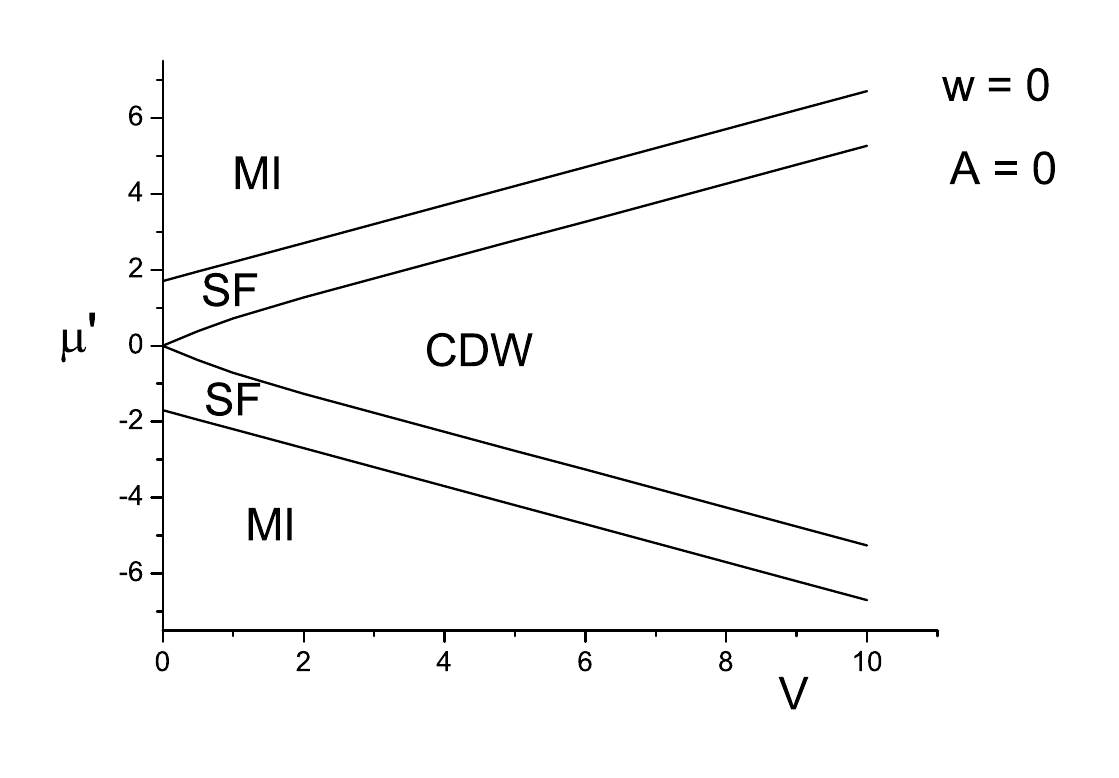}  
	\caption{ Phase diagrams for a one-dimensional proton
	conductor in the $(\mu', V)$ coordinates, $A = 0$, $(T = 0)$.}
	\label{fig6}
\end{figure}
% % % % % % % % % % % % %
\begin{figure}[!t]
	\centering
	\includegraphics[width=0.43\linewidth]{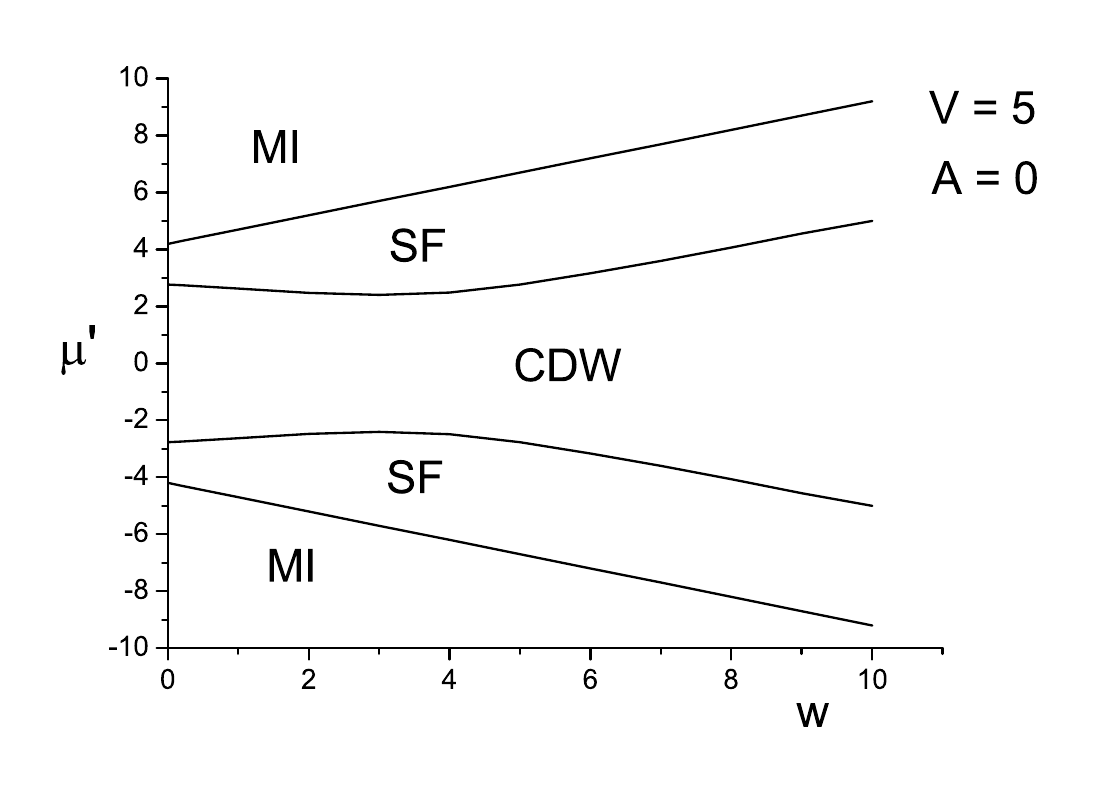} \qquad
	\includegraphics[width=0.43\linewidth]{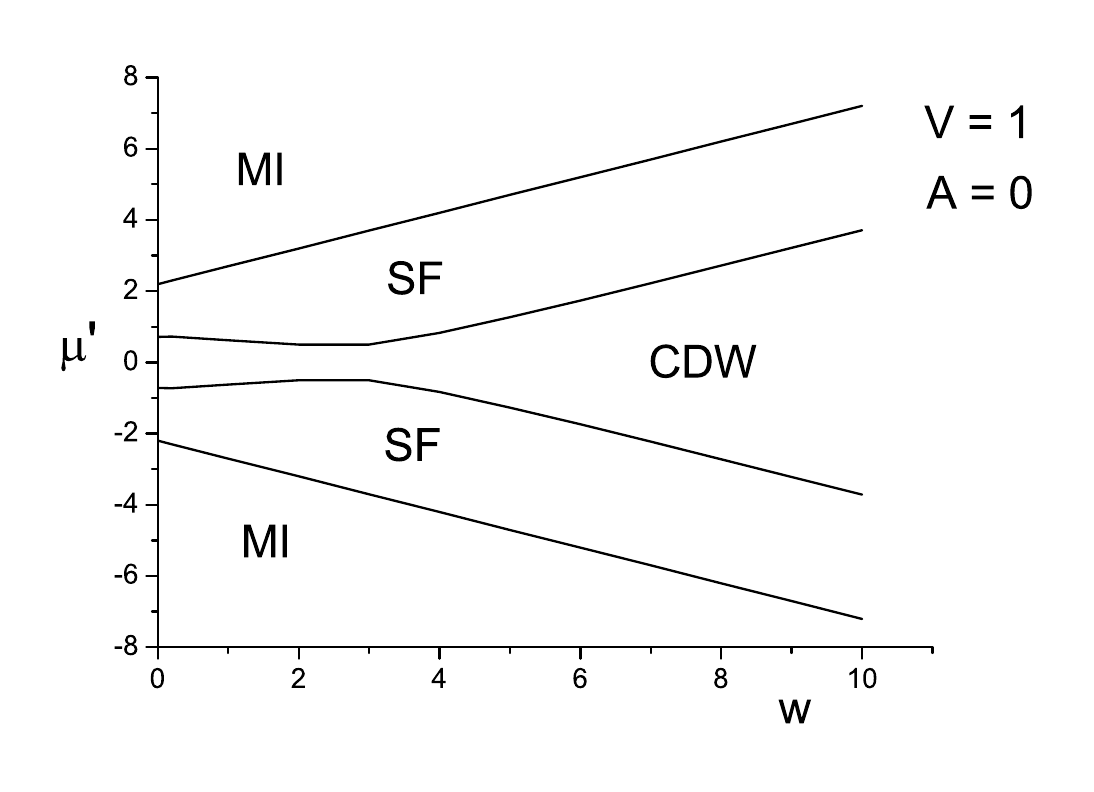} \\%\quad
	\includegraphics[width=0.43\linewidth]{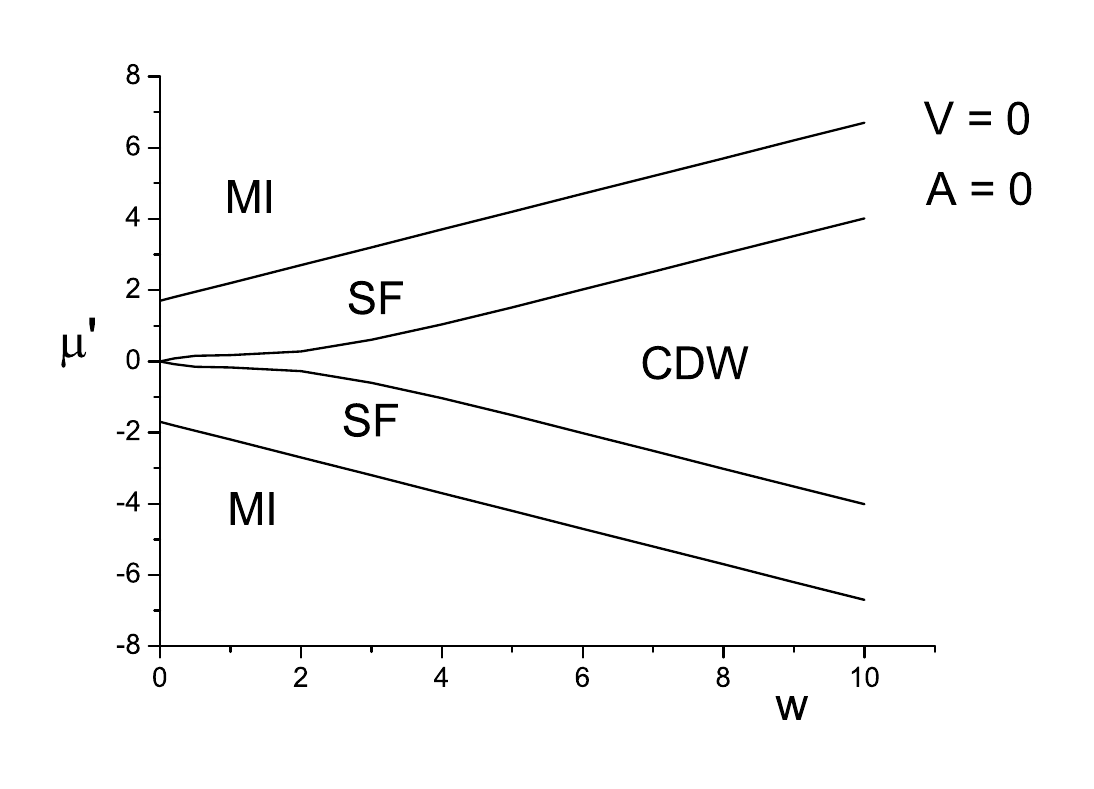}  
	\caption{Phase diagrams for a one-dimensional proton
	conductor in the ($\mu'$, \textit{w}) coordinates, $A = 0$, $(T = 0)$.}
	\label{fig7}
\end{figure}

The CDW, SF and  MI phases described above and the phase
transitions between them for a one-dimensional system exist only at
zero temperature. At small temperatures, we can distinguish the
regions of existence of the states of the CDW, SF and  MI
types as such in which the forms of the spectral functions of those
phases remain nearly the same as at $T = 0$. In this case, the
transition between the regions is not a genuine phase transition,
but has a crossover character.

Phase diagrams of equilibrium states of the system depending on the
short-range interactions between ions $V$ and \textit{w}, obtained at $T =
0, A = 0$, are shown in figure \ref{fig6} and figure \ref{fig7}.

 The characteristic features  of all the latter diagrams is that
the line separating SF and MI phases in ($\mu',V$) and
($\mu'$, \textit{w}) coordinates is strictly a straight line depending on $V$
or \textit{w}. At large values of interactions $V$ and \textit{w} and at their
further increase, we obtain the linear dependence of the width of
CDW phase on the values of interactions between particles. At
intermediate values of interactions there is a possibility of
narrowing of CDW phase and the presence of a minimum of its width
(in $\mu'$ coordinates) depending on the values of interactions
between ions.

\section{Conclusions}

The equilibrium states of one-dimensional proton conductors in the
systems with hydrogen bonds are investigated. Our extended hard-core
boson lattice model includes short-range interactions between ions
(parameter $V$ on a hydrogen bond, and parameter \textit{w} between
bonds), their transfer along hydrogen bonds with the two-minima
local anharmonic potential as well as their inter-bond hopping, and
the modulating field is taken into account. An exact diagonalization
method for a finite one-dimensional system with periodic boundary
conditions is used. The  existence of various phases of the system
at $T = 0$ depending on the values of interactions between particles
and the modulating field strength, is established by analyzing the
character of the obtained frequency dependence of one-particle
spectral density. It is shown that at $T = 0$, the repulsive
short-range interaction between particles ($V > 0$, $\textit{w} > 0$) results
in the emergence of a gap in the energy spectrum in the limit of
half-filling of ionic sites (the emergence of CDW phase). A
similar effect also takes place under the influence of the
modulating field $A$. Charge-density-wave (CDW) phase exists only
at half-filling of ionic sites ($\langle n_{a} \rangle + \langle
n_{b} \rangle)/2 = 1/2$. Departing from the half-filling, we get
into the SF phase. In this phase, the conductivity of the system
grows by a few orders. The mentioned phases and phase transitions
between them for the one-dimensional system exist only at $T = 0$.
The presence of SF phase is confirmed by the obtained sharp peak
of the real part of static susceptibility $\Re G_{k=0}(\omega = 0)$ at
$T = 0$. The predictable divergence of this susceptibility in the
SF phase $\Re G_{k=0}(\omega = 0) \rightarrow \infty$ was not
reached by us which is  predetermined by the finite size of the ion
conductor and probably by its onedimensionality. However, it is
shown that at small temperatures it is nearly by a few orders less
than at $T = 0$, and at the further increase of temperature a
maximum $\Re G_{k=0}(\omega = 0)$ becomes smeared  which confirms that
in onedimensional system a SF phase exists only at $T = 0$.

We have obtained phase diagrams of equilibrium states of the system
depending on the interactions between ions and the modulating field.
It is shown that the width of CDW phase (in $\mu'$ coordinates)
grows with an increase of the value of the modulating field. The
dependence of the width of CDW phase on short-range interactions
between ions $V$ and \textit{w} is more complex.  The characteristic
feature  of all the latter diagrams is that the line separating SF
and MI phases in ($\mu',V$) and ($\mu'$, \textit{w}) coordinates is strictly
a straight line depending on $V$ or \textit{w}. At large values of
interactions and at their further increase, we obtain a linear
dependence of the width of CDW phase on the values of interactions
between particles. At intermediate values of interactions, there is
a possibility of narrowing of CDW phase and the presence of a
minimum of its width (in $\mu'$ coordinates) depending on the values
of interactions between ions. We did not observe this kind of
behavior for the ion conductor (with the one-minimum local potential
for ions) (see \cite{Stetsiv}).

\newpage

\ukrainianpart

\title{Одночастинкові спектральні густини і фазові діаграми одновимірних протонних провідників}
\author{Р. Я. Стеців}
\address{Інститут фізики конденсованих систем Національної академії наук України, \\вул. Свєнціцького, 1, 79011 Львів, Україна}

\makeukrtitle

\begin{abstract}
\tolerance=3000%
Ми досліджуємо рівноважні стани одновимірних протонних провідників в
системах з водневими зв'язками. Наша  розширена модель жорстких
бозонів включає короткосяжну взаємодію між іонами водню, їх
перенесення як вздовж водневих зв'язків з двомінімумним локальним
ангармонічним потенціалом для протона, так і  перенос частинок між
сусідніми водневими зв'язками, а також модулююче поле. Розрахунки
проведено методом точної діагоналізації для скінчених одновимірних
систем з періодичними граничними умовами. Виходячи з характеру
частотної залежності отриманих одночастинкових спектральних густин,
встановлено існування різних фаз системи при $T = 0$ в залежності
від величини короткосяжної взаємодії між частинками і від величини
модулюючого поля; побудовані фазові діаграми.
\keywords протонний провідник, модель жорстких бозонів, спектральна
густина, фазові діаграми

\end{abstract}

\end{document}